\documentclass[a4paper]{article}

\usepackage{amsmath}
\usepackage{amsfonts}
\usepackage{amssymb}         
\usepackage{amsopn}
\usepackage{theorem}          
\usepackage{latexsym}
\usepackage{graphicx}        
\usepackage{mathrsfs}		
\usepackage{psfrag}
\usepackage{url}
\usepackage{float}
\usepackage{mathtools}
\usepackage{amsmath}
\usepackage{caption}
\usepackage{cite}
\usepackage{algpseudocode}
\usepackage{algorithm}
\usepackage{tikz}
\usepackage{xspace}
\algtext*{EndFor}
\algtext*{EndIf}
\usepackage[colorlinks=true,allcolors=blue,breaklinks,draft=false]{hyperref}   

\begin{document}

\thispagestyle{empty}
\begin{center} 

\title{A Survey of Shortest-Path Algorithms}

{\LARGE A Survey of Shortest-Path Algorithms} 
\par \bigskip
{\sc Amgad Madkour${}^{1}$, Walid G. Aref${}^1$, Faizan Ur Rehman${}^{2}$, Mohamed Abdur Rahman${}^2$, Saleh Basalamah${}^2$}   
\par \bigskip
\begin{minipage}{15cm}
\begin{flushleft}
{\small
\begin{itemize}
\item[${}^1$] {\it Purdue University, West Lafayette, USA} \\ 
\item[${}^2$] {\it Umm Al-Qura University, Makkah, KSA} \\ 
\end{itemize}
}
\end{flushleft}
\end{minipage}
\par \medskip \today
\end{center}
\par \bigskip

\begin{abstract} 

A shortest-path algorithm finds a path containing the minimal cost between two vertices in a graph. A plethora of shortest-path algorithms is studied in the literature that span across multiple disciplines. This paper presents a survey of shortest-path algorithms based on a taxonomy that is introduced in the paper. One dimension of this taxonomy is the various flavors of the shortest-path problem. There is no one general algorithm that is capable of solving all variants of the shortest-path problem due to the space and time complexities associated with each algorithm. Other important dimensions of the taxonomy include whether the shortest-path algorithm operates over a static or a dynamic graph, whether the shortest-path algorithm produces exact or approximate answers, and whether the objective of the shortest-path algorithm is to achieve time-dependence or is to only be goal directed. This survey studies and classifies shortest-path algorithms according to the proposed taxonomy. The survey also presents the challenges and proposed solutions associated with each category in the taxonomy.

\end{abstract}

\section{Introduction}
The shortest-path problem is one of the well-studied topics in computer science, specifically in graph theory. An optimal shortest-path is one with the minimum length criteria from a source to a destination. There has been a surge of research in shortest-path algorithms due to the problem's numerous and diverse applications. These applications include network routing protocols, route planning, traffic control, path finding in social networks, computer games, and transportation systems, to count a few. 

There are various graph types that shortest-path algorithms consider. A \textit{general graph} is a mathematical object consisting of vertices and edges. An  \textit{aspatial graph} contains vertices where their positions are not interpreted as locations in space. On the other hand, a \textit{spatial graph} contains vertices that have locations through the edge's end-points. A \textit{planar graph} is plotted in two dimensions with no edges crossing and with continuous edges that need not be straight.

There are also various settings in which a shortest-path can be identified. For example, the graph can be \textit{static}, where the vertices and the edges do not change over time. In contrast, a graph can be \textit{dynamic}, where vertices and edges can be introduced, updated or deleted over time. The graph contains either \textit{directed} or \textit{undirected} edges. The weights over the edges can either be \textit{negative} or \textit{non-negative} weights. The values can be real or integer numbers. This relies on the type of problem being issued.

The majority of shortest-path algorithms fall into two broad categories. The first category is \textit{single-source shortest-path} (SSSP), where the objective is to find the shortest-paths from a single-source vertex to all other vertices. The second category is \textit{all-pairs shortest-path} (APSP), where the objective is to find the shortest-paths between all pairs of vertices in a graph. The computation of shortest-path can generate either \textit{exact} or \textit{approximate} solutions. The choice of which algorithm to use depends on the characteristics of the graph and the required application. For example, approximate shortest-path algorithms objective is to produce fast answers even in the presence of a large input graph. A special sub-graph, called a \textit{spanner}, can also be created from the main graph that approximates the distances so that a shortest-path can be computed over that sub-graph.

\begin{figure}[H]
\centering
\captionsetup{type=figure}
\includegraphics[scale=0.8]{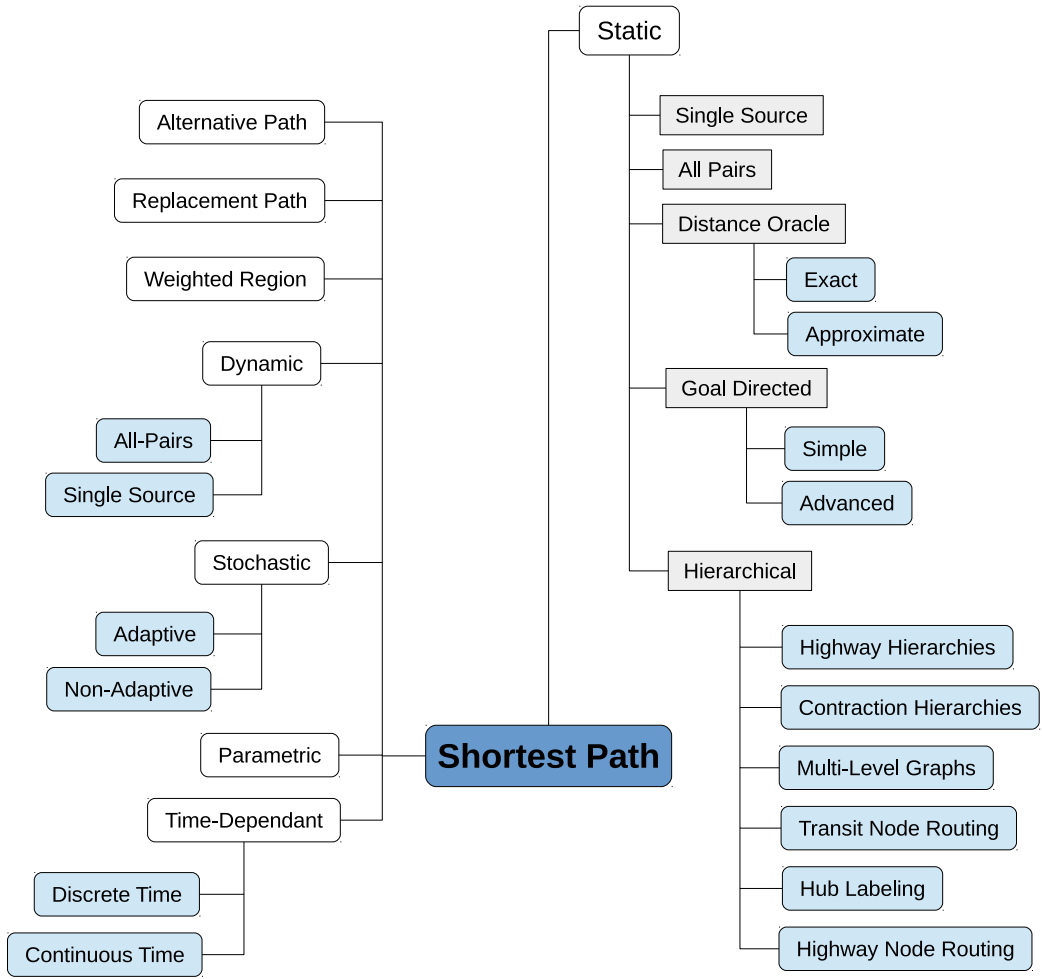}
\caption{Taxonomy of Shortest-Path Algorithms}
\label{fig:taxonomy}
\end{figure}

Given the large body of literature on algorithms for computing the shortest-path, the objective of this survey is to present a breakdown of these shortest-path algorithms through an appropriate taxonomy. The taxonomy aims to help researchers, practitioners, and application developers understand how each shortest-path algorithm works and to help them decide which type or category of shortest-path algorithms to use given a specific scenario or application domain. Figure~\ref{fig:taxonomy} illustrates the proposed taxonomy where each branch describes a specific category of shortest-path problem.

\section{Taxonomy}
As in Figure~\ref{fig:taxonomy}, the proposed taxonomy classifies the various shortest-path algorithms into multiple high-level branches. 

The static branch in Figure~\ref{fig:taxonomy} lists algorithms that operate over graphs with fixed weights for each edge. The weights can denote distance, travel time, cost, or any other weighting criteria. Given that the weights are fixed, some static algorithms perform precomputations over the graph.  The algorithms try to achieve a trade-off between the query time compared to the precomputation and storage requirements. Static algorithms consists of two classical algorithms for shortest-path fall under the two main categories~(1)~\textit{Single-source} shortest-path (SSSP), and~(2)~\textit{All-pairs} shortest-path (APSP). The SSSP algorithms compute the shortest-path from a given vertex to all other vertices. The APSP algorithms compute the shortest-paths between all pairs of vertices in the graph. \textit{Hierarchical} algorithms break the shortest-path problem into a linear complexity problem. 
This can lead to enhanced performance in computation by orders of magnitude.  \textit{Goal-directed} algorithms optimize in terms of distance or time toward the target solution. \textit{Distance oracle} algorithms include a preprocessing step  
to speed up the shortest-path query time. \textit{Distance oracle} algorithms can either be exact or approximate.
The dynamic branch in Figure~\ref{fig:taxonomy} lists algorithms that process \textit{update} or \textit{query} operations on a graph over time. The update operation can insert or delete edges from the graph, or update the edge weights. The query operation computes the distance between source and destination vertices.  Dynamic algorithms include both \textit{(APSP)} and \textit{(SSSP)} algorithms.~\textit{Time-dependent} algorithms target graphs that change over time in a predictable fashion. \textit{Stochastic} shortest-path algorithms capture the uncertainty associated with the edges by modeling them as random variables.~\textit{Parametric} shortest-path algorithms compute a solutions based on all values of a specific parameter.~\textit{Replacement path} algorithms computes a solution that avoids a specified edge, for every edge between the source vertex and the destination vertex. Replacement paths algorithms achieve good performance by reusing the computations of each edge it avoids. On the other hand, \textit{alternative path} algorithms also computes a shortest path between vertices that avoids a specified edge. The distinguishing factor between both categories is that replacement paths are not required to indicate a specific vertex or edge. On the other hand, alternative shortest-paths avoids the specified edge on the shortest-path. The weighted-regions problem finds the approximate shortest-path on weighted planar divisions.

\section{Related Work}

Zwick~\cite{Zwick2001} survey adopts a theoretical stand-point with regards to the exact and approximate shortest paths algorithms. Zwick's survey addresses single-source shortest-path (SSSP), all pairs shortest-path (APSP), spanners (a weighted graph variation), 
and distance oracles. The survey illustrates the various variations that each category adopts when handling negative and non-negative edge weights
as well as directed and undirected graphs. Sen~\cite{Sen2009} surveys approximate shortest-paths algorithms with a focus on spanners and distance oracles. Sen's survey discusses how spanners and distance oracles algorithms are constructed and their practical applicability over a static all-pairs shortest-paths setting. Sommer~\cite{Sommer2012} surveys query processing algorithms that trade-off the index size and the query time. Sommer's survey also introduce the transportation network class of algorithms, and include algorithms for general graphs as well as planar and complex graphs. 

Many surveys focus on algorithms that target traffic applications, especially route planning methods. In such related work, a network denotes a graph. Holzer et al.~\cite{Holzer2005} classify variations of Dijkstra's algorithm according to the adopted speedup approaches.
Their survey emphasizes on techniques that guarantee correctness. 
It argues that the effectiveness of speed-up techniques highly relies on the type of data.
In addition, the best speedup technique depends on the layout, memory and tolerable preprocessing time. In contrast to optimal shortest-path algorithms, Fu et al.~\cite{Fu2006} survey algorithms that target heuristic shortest-path algorithms  to quickly identify the shortest-path.  Heuristic algorithms aim is to minimize computation time. The survey proposes the main distinguishing features of heuristic algorithms as well as their computational costs. Goldberg~\cite{Goldberg2007} investigates the performance of point-to-point shortest-path algorithms over road networks from a theoretical standpoint. Goldberg reviews algorithms, e.g., Dijkstra and $A*$, and illustrates heuristic techniques for computing the shortest-path given a subset of the graph. The survey proves the good worst-case and average-case bounds over a graph. Also, it discusses reach-based pruning and illustrates how all-pairs shortest-path algorithms can be altered to compute reaches while maintaining the same time bound as their original counterparts. Delling and Wagner~\cite{Delling2009} survey route planning speedup techniques over some shortest-path problems including dynamic and time-dependent variants.  For example, the authors argue that shortcuts used in static networks cannot work in a time-dependent network. In essence, they investigate which networks can existing techniques be adopted to. Bast~\cite{Bast2009} illustrates speed-up techniques for fast routing between road networks and transportation networks. Bast's survey argues that the algorithms for both networks are different and require specialized speed-up techniques for each. Also, the survey presents how the speed-up
technique performs against Dijkstra's algorithm. Moreover, the survey presents two open questions, namely, (1)~how to achieve speed-up despite the lack of a hierarchy in transportation networks, and (2)~how to efficiently compute local searches, e.g., as in neighborhoods.

Demetrescu and Italiano~\cite{Demetrescu2006} survey algorithms that investigate fully dynamic directed graphs with emphasis on dynamic shortest-paths and dynamic transitive closures. The survey focuses on defining the algebraic and combinatorial properties as well as tools for dynamic techniques. The survey tackles two important questions, namely whether dynamic shortest-paths achieve a space complexity of $O(n^2)$, and whether single-source shortest path algorithms in a fully-dynamic setting be solved efficiently over general graphs. Nannicini and Liberti~\cite{Nannicini2008} survey techniques for dynamic graph weights and dynamic graph topology. They list classical and recent techniques for finding trees and shortest-paths in large graphs with dynamic weights. They target two versions of the problem, namely, time-dependence, and what they refer to as cost updates of the weights. Dean's survey~\cite{Dean2004} focuses on time-dependent techniques in a dynamic setting. It surveys one special case, namely, the First-In-First-Out (FIFO) network as it exposes structural properties that allow for the development of efficient polynomial-time algorithms.

This survey presents these aspects that are different from all its predecessors. First, it presents a taxonomy that can aid in identifying the appropriate algorithm to use given a specific setting. Second, for each branch of the taxonomy, the algorithms are presented in  chronological order that captures the evolution of the specific ideas and algorithms over time. Moreover, our survey is more comprehensive. We cover more recent algorithms that have been invented after the publication of the other surveys.

\section{Problem Definition}
Given a set of vertices $V$, a source vertex $s$, a destination vertex $d$, where $s,d \in V$, and a set of weighted edges $E$, over the set $V$, find the shortest-path between $s$ and $d$ that has the minimum weight. The input to the shortest-path algorithm is a graph $G$ that consists of a set of vertices $V$ and edges $E$. The graph is defined as $G=(V,E)$. The edges can be \textit{directed}  or \textit{undirected}.  The edges have explicit weights, where a weight is defined as $w(e)$, where $e \in E$, or unweighted, where the implicit weight is considered to be 1. When calculating the algorithm complexity, we refer to the size of the set of vertices $V$ as $n$ and the size of the set of edges $E$ as $m$.

%

\section{Static Shortest-Path Algorithms}
In this section, we review 
algorithms for both the single-source shortest-path (SSSP) and all-pairs shortest-path (APSP) problems.

\subsection{Single-Source Shortest-Path (SSSP)}

\textit{Definition:} Given a Graph $G=(V,E)$ and Source $s \in V$, compute all distances $\delta(s,v)$, where $v \in V$.

The simplest case for SSSP is when the graph is unweighted. Cormen et al.~\cite{Cormen2001}~suggest that breadth-first search can be simply employed by starting a scan from  a root vertex and inspecting all the neighboring vertices. For each neighboring vertex, it probes the non-visited vertices until the path with the minimum number of edges from the source to the destination vertex is identified.

Dijkstra's algorithm~\cite{Dijkstra1959} solves the single source shortest-path (SSSP) problem from a given vertex to all other vertices in a graph. Dijkstra's algorithm is used over directed graphs with non-negative weights. The algorithm identifies two types of vertices:~(1)~Solved and (2)~Unsolved vertices. It initially sets the source vertex as a solved vertex and checks all the other edges (through unsolved vertices) connected to the source vertex for shortest-paths to the destination. Once the algorithm identifies the shortest edge, it adds the corresponding vertex to the list of solved vertices. The algorithm iterates until all vertices are solved.
Dijkstra's algorithm achieves a time complexity of $O(n^{2})$. One advantage of the algorithm is that it does not need to investigate all edges. This is particularly useful when the weights on some of the edges are expensive. The disadvantage is that the algorithm deals only with non-negative weighted edges. Also, it applies only to static graphs. 

Dijkstra's algorithm performs a brute-force search in order to find the optimum shortest-path and as such is known to be a greedy algorithm. Dijkstra's algorithm follows a successive approximation procedure based on Bellman Ford's optimality principle~\cite{Bellman1957}. This implies that Dijkstra's algorithm can solve the dynamic programming equation through a method called the reaching method~\cite{Denardo2003,Sniedovich2006,Sniedovich2010}. The advantage of dynamic programming is that it avoids the brute-force search process by tackling the sub-problems. Dynamic programming algorithms probe an exponentially large set of solutions but avoids examining explicitly all possible solutions. The greedy and the dynamic programming versions of Dijkstra's algorithm are the same in terms of finding the optimal solution. However, the difference is that both may get different paths to the optimal solutions.

Fredman and Tarjan~\cite{Fredman1987}~improve over Dijkstra's algorithm by using a Fibonnaci heap (F-heap). This implementation achieves $O(nlogn + m)$ running time because the total incurred time for the heap operations is $O(n~log~n+m)$ and the other operations cost $O(n+m)$. Fredman and Willard~\cite{Fredman1990a,Fredman1990,Fredman1993} introduce an extension that includes an $O(m+n~log~n/loglog~n)$ variant of Dijkstra's algorithm through a structure termed the AF-Heap. The AF-Heap provides constant amortized costs for most heap operations and $O(log~n/loglog~n)$ amortized cost for deletion. Driscoll and Gabow~\cite{Driscoll1988} propose a heap termed the relaxed Fibonacci heap. A relaxed heap is a binomial queue that allows heap order to be violated.  The algorithm provides a parallel implementation of Dijkstra's algorithm. 

Another line of optimization is through improved priority queue implementations. Boas~\cite{Boas1975} and Boas et al.~\cite{Boas1976} implementations are based on a stratified binary tree. The proposed algorithm enables online manipulation of a priority queue. The algorithm has a processing time complexity of $O(loglog~n)$ 
and storage complexity of $O(n~loglog~n)$. A study by Thorup~\cite{Thorup1996} indicates the presence of an analogy between sorting and the SSSP problem, where SSSP is no harder than sorting edge weights. Thorup~\cite{Thorup1996} describes a priority queue giving a complexity of $O(loglog~n)$ per operation and $O(m~loglog~n)$ complexity for the SSSP problem. The study examines the complexity of using a priority queue given memory with arbitrary word size. Following the same analogy,  Han~\cite{Han2001} proposes a deterministic integer sorting algorithm in linear space that achieves a time complexity of $O(m~loglog~n~logloglog~n)$ for the SSSP problem. The approach by Han~\cite{Han2001} illustrates that sorting arbitrarily large numbers can be performed by sorting on very small integers.

Thorup~\cite{Thorup1999}~proposes a deterministic linear space and time algorithm by building a hierarchical bucketing structure that avoids the sorting operation. A bucketing structure is a dynamic set into which an element can be inserted or deleted. The elements from the buckets can be picked in an unspecified manner as in a doubly-linked list. The algorithm by Thorup~\cite{Thorup1999} works by traversing a component tree. Hagerup~\cite{Hagerup2000}~improves over the algorithm of Thorup, achieving a time complexity of  $O(n+m~log~w)$, where $w$ is the width of the machine word. This is done through a deterministic linear time and space algorithm.

Bellman, Ford, and Moore~\cite{Bellman1958,Ford1956,Moore1957}~develop an SSSP algorithm that is capable of handling negative weights unlike Dijkstra's algorithm. It operates in a similar manner to Dijkstra's, where it attempts to compute the shortest-path but instead of selecting the shortest distance neighbor edges with shortest distance, it selects all the neighbor edges. Then, it proceeds in $n-1$ cycles in order to guarantee that all changes have been propagated through the graph. While it provides a faster solution than Bellman-Ford's algorithm, Dijkstra's algorithm is unable to detect negative cycles or operate with negative weights. However, if there is a negative cycle, then there is no shortest-path that can be computed. The reason is due to the lower total weight incurred due to the traversal cycle. Bellman-Ford's algorithm achieves a run-time complexity of $O(nm)$. Its strong points include the ability to operate on negative weights and detect negative cycles. However, the disadvantages include its slower run-time when compared to Dijkstra's algorithm. Also, Bellman-Ford's algorithm does not terminate when the iterations do not affect the graph weights any further. 

Karp~\cite{Karp1978} addresses the issue of whether a graph contains a negative cycle or not. He defines a concept termed minimum cycle mean and indicates that finding the minimum cycle mean is similar to finding the negative cycle. Karp's algorithm achieves a time complexity of $O(nm)$.

Yen~\cite{Yen1970} proposes two performance modifications over Bellman Ford, and Moore~ \cite{Bellman1958,Ford1956,Moore1957}. The first involves the relaxation of edges. An edge is relaxed if the value of the vertex has changes. The second modification is dividing the edges based on a linear ordering over all vertices. Then, the set of edges are partitioned into one or more subsets. This is followed by performing comparisons between the two sets according to the proposed partitioning scheme. A slight improvement to what Yen~\cite{Yen1970} proposes has been introduced by Bannister and Eppstein~\cite{Bannister2011} where instead of using an arbitrary linear ordering, they use a random ordering. The result is fewer number of iterations over both subsets.

\subsection{All-Pairs Shortest-Path (APSP)}

\textit{Definition:} Given a graph $G=(V,E)$, compute all distances between a source vertex $s$ and a destination v, where $s$ and $v$ are elements of the set $V$.

The most general case of APSP is a graph with non-negative edge weights. In this case, Dijkstra's algorithm can be computed separately for each vertex in the graph. The time complexity will be $O(mn+n^2logn)$~\cite{Karger}. 

A vast number of algorithms has been proposed that handle real edge-weights for the all-pairs shortest-path problem. Floyd-Warshall algorithm~\cite{Floyd1962,Warshall1962} tries to find all pairs shortest-paths (APSP) in a weighted graph containing positive and negative weighted edges. Their algorithm can detect the existence of negative-weight cycles but it does not resolve these cycles. The complexity of Floyd-Warshall algorithm is $O(n^3)$, where n is the number of vertices. The detection of negative-weight cycle is done by probing the diagonal path matrix. Floyd-Warshall algorithm cannot find the exact shortest-paths between vertices pairs because it does not store the intermediate vertices while calculating. However, using a simple update, one can store this information within the algorithm steps. The space complexity of the algorithm is $O(n^{3})$. However, this space complexity can reach $O(n^{2})$ by using a single displacement array. The strong point of the algorithm is that it can handle negative-weight edges and can detect negative-weight cycles. The main drawback though is that the timing complexity for running Dijkstra's algorithm on all vertices (to convert it from SSSP to APSP) will be $O(mn + n^{2} log n)$. This timing complexity is lower than $O(n^3)$ if and only if $m < n^{2}$ (i.e., having a sparse graph). 

Many studies have been proposed better running time over Floyd-Warshall's algorithm on~\textit{ real-valued edge weights}. A notable enhancement has been proposed by Fredman~\cite{Fredman1976} that relies on a matrix-oriented approach. His approach relies on the theorem proposed by Aho and Hopcroft~\cite{Aho1974} the complexity of an $N$x$N$ matrix multiplication using a min/plus multiplication approach is similar to that of shortest-paths. He shows that $O(N^{5/2})$ comparisons suffices to solve the all-pairs shortest-paths (APSP) problem. The algorithm achieves a complexity of $O(n^3 (log log n)/ log n^{1/3})$. Table~\ref{table:realvalue} summarizes the enhancements proposed for real-valued edges up to this date.

\begin{center}
\captionsetup{type=table}
\captionof{table}{Algorithms and complexities for real-valued edges}
\begin{tabular}{|c|c|}
\hline  Time Complexity & Author\\
\hline  $n^3$ &  \cite{Floyd1962,Warshall1962}\\ 
\hline  $n^3 (log log n)/ log n^{1/3}$ & \cite{Fredman1976}\\ 
\hline  $n^3 (log log n / log n)^{1/2}$ &  \cite{Takaoka1992}\\ 
\hline  $n^3 / (log n)^{1/2}$ & \cite{Dobosiewicz1990}\\ 
\hline  $n^3 (log log n / log n)^{5/7}$ &  \cite{Han2004}\\ 
\hline  $n^3 log log n / log n$ &  \cite{Takaoka2004}\\ 
\hline  $n^3 (log log n)^{1/2} / log n$ &  \cite{Zwick2004}\\ 
\hline  $n^3/log n)$ &  \cite{Chan2006}\\ 
\hline  $n^3 (log log n / log n)^{5/4}$ &  \cite{Han2006}\\ 
\hline  $n^3 (log log n)^3 / (log n)^2$ &  \cite{Chan2007}\\ 
\hline  $n^3 (log log n)/ (log n)^2$ & \cite{Han2012}\\
\hline
\end{tabular} 
\label{table:realvalue}
\end{center}

The best result by Han and Takaoka~\cite{Han2012} achieve $O(log log n)^2$ reduction factor when compared to the result of~\cite{Chan2007}. Their approach focuses on the distance product computation. First, an $n$x$n$ matrix js divided into $m$ sub-matrices, each having $n$x$n$/$m$ dimensions, where m is determined based on a specific criterion. Then, the algorithm proceeds in a series of matrix manipulations, index building, encoding, and partitioning steps until it reaches the proposed bound.

The best~\textit{non-negative edge weight} complexity is $O(n^2 log n)$~ \cite{Moffat1987}. First, the algorithm sorts all adjacency lists in an increasing weight fashion. Then, it performs an SSSP computation $n$ times and proceeds in iterations. In the first phase, it uses the notion of \textit{potential} over the edges of vertices and selects and labels the edge with the minimum potential. \textit{Potential} derived from the \textit{potential-model} is defined as a probability distribution on complete directed graphs with arbitrary edge lengths that contain no negative cycles. The algorithm runs in two main phases, each with a specific invariant and has an $O(n^2 log n)$ complexity.

The best~\textit{positive integer edge weight} complexity is $O(n^{\omega}+c)$~\cite{Roditty2011},  where $\omega < 2.575$ is the exponent being proposed by Coppersmith and Winograd~\cite{Coppersmith1990}. Their proposed algorithm provides a transition between the fastest exact and approximate shortest-paths algorithms with a linear error rate. The algorithm focuses on directed graphs with small positive integer weights in order to obtain additive approximations. The approximations are polynomial given the actual distance between pairs of vertices.

\subsection{Distance Oracles}

\textit{Definition:} Given a graph $G=(V,E)$, a distance oracle encompasses a~(1)~data structure or index that undergoes preprocessing, and a~(2)~query algorithm.

The term \textit{distance oracle} has been proposed by Thorup and Zwick~\cite{Thorup2005}. It proposes a faster alternative to the SSSP and APSP algorithms. This can be achieved by preprocessing the graph and creating an auxiliary data structure to answer queries. Distance oracle operates in two phases, namely, a preprocessing phase and a query phase. In the preprocessing phase, information such as data structures or indexes are computed. In contrast, the query processing phase processes queries efficiently using the outcome from the preprocessing phase. Distance oracles may return exact or approximate distances. A distance oracle provides an efficient trade-off between space (in terms of data structure or index storage) and query time.

\subsubsection{Exact Distances}

Fakcharoenphol and Rao~\cite{Fakcharoenphol2006} propose an algorithm for planar graphs that balances the trade-off between preprocessing and query time. The preprocessing complexity for both space and time is ~$\tilde{O}(n)$, and the run-time complexity is $\tilde{O}(\sqrt{n})$. Their proposed approach creates a non-planar graph given a subset of vertices followed by the computation of the shortest-path tree. First, the graph is divided into a set of bipartite graphs. The distance matrices of the bipartite graph need to comply with a non-crossing condition referred to as the Monge condition. The proposed result of $O(\sqrt{n})$ holds as long as the non-crossing condition is enforced.

Klein et al.~\cite{Klein2010} propose a linear-space algorithm with a fast preprocessing complexity of $O(nlog^{2} n)$, over a directed planar graph. The graph can include both positive and negative edges. 

Given a planar directed graph $G$, and a source vertex, the algorithm finds a curve known as a Jordan curve. A Jordan curve $C$ is identified if it passes through $O(\sqrt{n})$ vertices. A \textit{boundary vertex} is one that passes through $C$. Cutting the graph and duplicating the boundary vertices creates subgraphs $G_i$. The algorithm passes through five stages:~(1)~recursively compute the distances from $r$ within a graph where $r$ is an arbitrary boundary vertex,~(2)~compute all distances between boundary vertices,~(3)~use a variant of Bellman-Ford to compute the graph distances from the boundary vertex $r$ to all other boundary vertices,~(4)~use Dijkstra's algorithm to compute the graph distances from the boundary vertex $r$ to all other vertices,~(5)~use Dijkstra's algorithm to compute graph distances from the source Vertix. This requires time of $O(nlog n)$.

Djidjev~\cite{Djidjev1996} proposes a faster query time algorithm and proves that for any $S \in [n,n^2]$, a distance oracle can have a space complexity during preprocessing of $O(S)$, and query time complexity of $O(n^2/S)$. Djidjev's objective is to have an algorithm in which the product of preprocessing-space and query-time is not greater than those of SSSP and APSP problems. The proposed algorithm provides a complexity of $O(\sqrt{n})$ for any class of directed graphs where the separator theorem holds.

Cabello~\cite{Cabello2006} improves the preprocessing time, and provides a theoretical proof that, for any $S \in [n^{4/3},n^2]$, a distance oracle can have $O(S)$ preprocessing space complexity, and $O(n/\sqrt{S})$ query time  complexity. This is slower than the algorithm proposed by Djidjev~\cite{Djidjev1996} by a logarithmic factor but still covers a wider range of $S$. The proposed approach constructs a data structure between any pair of vertices that can answer distance-based queries. Then, the algorithm queries the data structure with those pairs.

Wulff-Nilsen~\cite{Wulff-Nilsen2013} proposes a constant query-time algorithm for unweighted graphs, and proves that for any $S \in [(log n/loglog n)^2,n^{2/5}]$, a distance oracle can have a space complexity of $o(n^2)$. The algorithm relies on the Wiener index of a graph. The Weiner index defines the sum of distances between all pairs of vertices in a graph. The proposed technique shows the existence of subquadratic time algorithms for computing the Wiener index. Computing the Wiener index has the same complexity as computing the average vertex pairs distances.

Henzinger et al.~\cite{Henzinger1997} propose a SSSP algorithm requiring $O(n^{4/3} log(nL))$ time, where $L$ is the absolute value of an edge with the smallest negative value. The proposed algorithm also achieves a similar bound for planar graphs and planar bipartite graphs. They also propose a parallel and dynamic variant of the algorithm. The key component of their approach is the use of graph-decompositions based on planar separators. 

Mozes and Sommer~\cite{Mozes2012} propose an algorithm to answer distance queries between pairs of vertices in planar graphs with non-negative edge weights. They prove that, for any $S \in [n loglog n, n^2]$, a distance oracle can have $\tilde{O}(S)$ preprocessing time complexity, and $O(S)$ space complexity. Distance queries can be answered in $\tilde{O}(n/\sqrt{S})$. The graph can be preprocessed in $\tilde{O}(n)$ and the generated data structure will have a size of $O(n loglog c)$. The query time will be $\tilde{O}(c)$ where $C$ is a cycle with $c = O(\sqrt{n})$ vertices.

\subsubsection{Approximate Distances}

Approximate distance oracles algorithms attempt to compute shortest-paths by querying only some of the distances. It is important to note that algorithms that deal with finite metric spaces produce only approximate answers. Some algorithms create \textit{spanners}, where a spanner is a sparse sub-graph  that approximates the original graph. They can be regarded as a spanning tree that maintains the locality aspects of the graph. These locality aspects defines a \textit{stretch} where a stretch is a multiplicative factor that indicates the amount distances increase in the graph. The stretch is a result of utilizing the spanner edges only~\cite{Elkin2004}.

Other algorithms approximate distances by triangulation using a concept called~\textit{landmark} or~\textit{beacon}~\cite{Sommer2012} that is selected by random sampling, where each vertex stores distances to all landmarks. Note that given the definition of approximate distance oracles, the actual shortest-path is still not guaranteed to be retrieved.

Zwick~\cite{Zwick1998} presents an APSP algorithm for directed graphs that utilizes a matrix multiplication where the approximate distance is computed in $O((n^\omega/\epsilon)log(W/\epsilon)$, where $\epsilon > 0$ for any $\epsilon$. They define the stretch as $1+\epsilon$ and $W$ represents the largest weighted edge identified in the graph.

Aingworth et al.~\cite{Aingworth1996} propose an APSP algorithm for undirected graphs with unweighted edges that does not adopt a matrix multiplication approach. A trade-off of not using fast matrix multiplication is a small additive error. They propose two algorithms; one that achieves an additive error of 2 in time $O(n^{2.5}\sqrt{log~n})$. They also provide an estimate of graph paths and distances in $O(n^{5/2}(log~n)^{1/2})$ and another 2/3-approximation algorithm that achieves a query time of $O(m(n~log~n)^{1/2})$.

Dor et al.~\cite{Dor2000} improve on previous surplus results by proposing an APSP algorithm that computes the surplus 2 estimate in $\tilde{O}(n^{3/2} m^{1/2})$. They also show that, for any $k$, a surplus 2(k-1) estimate takes $\tilde{O}(kn^{2-1/k} m^{1/k})$ to be computed. Their work relies on the one main observation that there is a set of vertices that represent vertices with high degree value. In other words, a set of vertices $X$ is said to represent a set of $Y$ if all vertices in $X$ have a neighbor in $Y$.

Cohen and Zwick~\cite{Cohen2001} improve the work proposed by Dor et al.~\cite{Dor2000} for weighted undirected graphs by proposing an algorithm that computes the surplus 2 estimate of all distances in $\tilde{O}(n^{3/2}m^{1/2})$ and 3 estimate in $\tilde{O}(n^2)$. They show that finding the estimated distances between all-pairs in directed graphs is a hard problem, similar to the Boolean Matrix multiplication. This makes their proposed approximation algorithm only valid for undirected graphs. Their algorithm relies on two important aspects: partitioning of the graph with the assumption that it is directed and the use of an SSSP algorithm, e.g., Dijkstra's.

Patrascu and Roditty~\cite{Patrascu2010} further improve the stretch bound of intermediate vertices on the expense of increasing the space requirements and achieve $\tilde{O}(n^{2/3})$. This approach defines the notion of \textit{balls}, defined as $B$, where balls around each vertex grow geometrically and stop based on a specific criteria. Given the vertices $s$ and $t$, the worst-case happens when the balls do not intersect.

Agarwal et al.~\cite{Agarwal2011} also propose a 2 estimate approach that can be implemented in a distributed fashion. The approach is mainly meant for compact routing protocols. It aims to characterize the space and time trade-off for approximate distance queries in sparse graphs. For both approaches above (i.e.,~\cite{Patrascu2010} and \cite{Agarwal2011}), the space versus query time trade-off depends on the number of edges.

For spanners, Elkin and Peleg~\cite{Elkin2004} propose a general $(1+\epsilon,\beta)$-spanner with space complexity of $O(\beta n^{1+1/k})$, where $\beta = \beta(\kappa,\epsilon)$ is a constant when $\kappa$ and $\epsilon$ are also constants. They claim that the stretch and spanners can be minimized in a simultaneous evaluation fashion.

Baswana and Sen~\cite{Baswana2007} propose a spanner with a $(2k-1)$ stretch that can be computed in $O(km)$ and with a size of $O(kn^{1+1/k})$, where $k > 1$. They provide a theoretical proof that a spanner with a $(2k-1)$ stretch can be computed without distance computation in linear time through a novel clustering technique. The proposed approach can take $O(k)$ rounds. Each round explores an adjacency vertex list in order to determine the edges that need to be removed. The advantage of this approach is its applicability to various computational environments, e.g., the synchronous distributed model, the external memory model, and the CRCW PRAM model.

For planar graphs, Thorup~\cite{Thorup2004} proposes an $(1+\epsilon)$-approximate distance oracle. This approach provides a constant number of shortest-paths through separators in contrast to Lipton et al.~\cite{Lipton1979}. For each vertex, it stores the shortest-path distances to a set of $O(1/\epsilon)$ landmarks per level. This process is performed recursively for $O(log n)$ levels.

Kawarabayashi et al.~\cite{Kawarabayashi2011} propose a planar graph algorithm that provides tunable trade-offs, where a polylogarithmic query time can be achieved while maintaining a linear space requirement with respect to the graph size. The proposed approach achieves a preprocessing time complexity of $O(n log^2 n)$ and query time of $O(\epsilon^{-2} log^2 n)$. It achieves faster running time than Thorup's approach that computes a set $C$ of connections that covers all vertices of a graph with every vertex containing $O(\epsilon^{-1})$ connections~\cite{Thorup2004}. In contrast, only a subset of vertices is covered using Kawarabayashi et al. approach. The approach is $O(\epsilon^{-1})$ times the number of paths in space complexity.

For complex networks, Chen et al.~\cite{Chen2012} proposes a distance oracle over random power-law graphs~\cite{Aiello2000} with 3 estimate that has a space complexity of $O(n^{4/3})$.  Their approach adopts the distance oracle proposed by Thorup and Zwick~\cite{Thorup2005}, where they use high-degree vertices as landmarks. The adaptation includes selecting vertices with the highest degree as landmarks. It encodes the shortest-paths in the vertex labels.

\subsection{Goal-Directed Shortest-Paths}

A goal-directed shortest-path search algorithm is based on adding annotations to vertices or edges of the graph that consist of additional information. This information allows the algorithm to determine which part of the graph to prune in the search space.

\subsubsection{Simple Goal-Directed Search}

Hart et al.~\cite{Hart1968} propose a simple goal-directed algorithm, termed $A^{*}$. The algorithm proposes a heuristic approach in finding the shortest-path. Unlike Dijkstra's algorithm, $A^{*}$ is an informed algorithm, where it searches  the routes that lead to the $A^{*}$ final goal. $A^{*}$ is an optimal best-first-search greedy algorithm. But what sets $A^{*}$ aside from other algorithms is its ability to maintain the distance it traveled into account. $A^{*}$ always finds the shortest-path if an admissible heuristic function is used. The strong point of the algorithm is that it is meant to be faster than Dijkstra since it explores less number of vertices. On the downside, if $A^{*}$ does not use a good heuristic method, it will not reach the shortest-path.

Some of the variants of the $A^{*}$ algorithm use landmarks and other techniques in order to achieve better performance than $A^{*}$ under various setups. Goldberg and Werneck~\cite{Goldberg2005} propose a preprocessing phase where initially a number of landmarks are selected followed by the computation of the shortest-path where it is stored between the vertices of all these landmarks. They propose a constant-time lower-bound technique using the computed distances in addition to the triangle inequality property. The lower-bound technique is based on the $A^{*}$ algorithm, the landmark chosen, and the triangle inequality. 

Gutman~\cite{Gutman2004} offers a comparable solution to the problem, where his work is based on the concept of reach. Gutman's technique relies on storing a reach value and the Euclidean coordinates of all vertices. The advantage of Gutman's approach is that it can be combined with the $A^{*}$ algorithm when compared to the work by Goldberg and Werneck~\cite{Goldberg2005}, Gutman's~\cite{Gutman2004} outperforms their proposed technique given one landmark while it performs worse given sixteen landmarks. On the downside, Gutman's approach depends on domain-specific assumptions, longer preprocessing complexity, and inapplicability in a dynamic setting.

Potamias et al.~\cite{Potamias2009} propose an approximate \textit{landmark-based} technique for point-to-point distance estimation over large networks. A theoretical proof is presented to indicate that the problem is NP-Hard and they propose heuristic solutions. In specific, they propose a smart landmark selection technique that can yield higher accuracy, reaching 250 times less space than selecting landmarks at random. Among their evaluated strategies, the Centrality is more robust than the Degree strategy. Also, strategies based on partitioning, e.g., Border/P exhibit better computational cost across datasets.

Kleinberg et al.~\cite{Kleinberg2004} propose an algorithm with provable performance guarantees for beacon-based triangulation and embedding. The beacon-based algorithms are basically designed for triangulation, where they use the triangle inequality to deduce the unmeasured distances. They indicate that a multiplicative error of $1+\delta$ on a $1-\epsilon$ fraction of distances can be achieved by triangulation-based reconstruction given a constant number of beacons. The algorithm also achieves a constant distortion over $1-\epsilon$ of distances.

Maue2009 et al.~\cite{Maue2009} claim that Dijkstra's algorithm can be enhanced by precomputing the shortest-path distances. They propose to partition the graph into $k$ non-overlapping clusters and perform two operations; (1) store the start and end point, (2) store the shortest connection between each pair of clusters. The proposed algorithm achieves a speed-up scaling factor of $\sqrt{k}$ in contrast to Dijkstra's algorithm.

\subsubsection{Advanced Goal-Directed Search}

Edge labels is an approach that relies on precomputing the information for an edge $e$ and vertices $M$. The superset $M(e)$ represents all the vertices on a shortest-path that start with an edge $e$. The graph is first partitioned into a set of regions of the same size alongside a precomputed set of boundary vertices. In order to compute the edge flags, an SSSP computation is done on the regions for all the boundary vertices. Various work, e.g., Kohler et al.~\cite{Kohler2005}, Schulz et al.\cite{Schulz1999}, and Lauther~\cite{Lauther2004} further present some of the edge-label variations.

M\"{o}hring et al.~\cite{Mohring2007} propose an algorithm for sparse directed graphs with non-negative edge weights, termed the~\textit{arc-flag} approach. The arc-flag approach preprocesses graph data to generate information that speeds up shortest-path queries by dividing the graph into regions and determining if an arc in a specific region lies on the shortest-path. Given a suitable partitioning scheme and a bi-directed search, the arc-flag approach 500 times faster than the standard Dijkstra's algorithm over a large graph. Schilling et al.~\cite{Schilling2006} present a further improvement by searching once for each region. Their approach achieves speed-up of more than 1,470 on a subnetwork of 1 million vertices. 

Goldberg and Werneck~\cite{Goldberg2005} propose an~$A^{*}$~based search Landmarks (ALT) algorithm that uses the triangle inequality. They show that precomputing the distances to a set of landmarks can bound the shortest path computational cost. They propose an average of 20 landmarks that are well-distributed over the corners of the graph. In turn, their approach leads to speed up for route planning.
 
Bauer et al.~\cite{Bauer2010} study how to systematically combine speed-up techniques proposed for Dijkstra's algorithm, e.g., adding goal-direction approaches to hierarchical approaches. They present generalized technique that demonstrates how speed-up performance can be improved. Their results show that Highway vertex Routing and Arc-Flags achieves the best speed-up while maintaining an adequate preprocessing cost. They also present a hierarchical $A^{*}$-based search Landmarks (ALT) algorithm on dense graphs.

Delling et al.~\cite{Delling2012} present an algorithm termed round-based public transit router (RAPTOR). RAPTOR is not based on Dijkstra's algorithm as it probes each route in the graph at most once. RAPTOR works in fully dynamic scenarios and can be extended to handle, for example, flexible departure times.

Bauer and Delling~\cite{Bauer2009} uses hierarchical based techniques to extend the edge flag approach, e.g., using contraction hierarchies during preprocessing, and hence tackling a main processing drawback of edge flags. The proposed work is termed (Shortcuts + Arc-Flags) or SHARC, for short. The key observation about SHARC is that it is enough to set sub-optimal edge flags to most edges, and this focuses the preprocessing on important edges only. Another observation is that SHARC incorporates hierarchical aspects implicitly. SHARC also extends the edge flag approach of M\"{o}hring et al.~\cite{Mohring2007} to achieve a fast unidirectional query algorithm.

Maue et al.~\cite{Maue2009} propose a goal-directed algorithm that utilizes precomputed cluster distances (PCD). The proposed approach first partitions the graph into clusters. This is followed by precomputing the shortest connections between the pairs of clusters $U$ and $V$. PCDs produce bounding factors for distances that can be used to prune the search when compared with the $A^{*}$ algorithm. In turn, this achieves a speed-up comparable to ALT while using less space.

\subsection{Hierarchical Shortest-Path}

Hierarchical shortest-path algorithms deal with generating a multi-layered vertex hierarchy in the preprocessing stage. A hierarchical structure is prominent in areas, e.g., road networks, where it exhibits hierarchical properties, e.g., ordering important streets, motorways, and urban streets~\cite{Schultes2005}. 

In general, methods using contraction hierarchies provide low space complexity. Contraction hierarchies contain many variants such as reach-based methods and highway hierarchies and vertex routing. On the other hand, Transit-vertex Routing and Hub Labels provide fast query-time~\cite{Sommer2012}.

The following sections discuss various algorithms that follow a hierarchical approach.

\subsubsection{Highway Hierarchies}

Highway Hierarchies capture edge-based properties. For example, highway edges exhibit a better representation for shortest paths although they may not be located between the source and the destination vertices. The algorithm generates a hierarchy of graphs that enables fast query time with correctness guarantees.

Sanders and Schultes~\cite{Sanders2005,Sanders2006} propose a static undirected highway hierarchies algorithm around the notion of correctly defining local search and highway network appropriately. They define local search  as one that visits $H$ (tuning parameter) closest vertices from the source or target. A highway edge is created if it lies on the path from the source vertex to the destination vertex with that edge not being within the $H$ closest vertices from the source or destination.

Nannixini et al.~\cite{Nannicini2010} propose an algorithm that relies on time-dependent lengths. They extend the original algorithm by Sanders and Schultes~\cite{Sanders2005} to the case of directed graphs. Their aim is to find the fastest paths on a large dynamic road network that have quasi real-time updates.

\subsubsection{Contraction Hierarchies}

A contraction hierarchy has a level for each vertex reaching up to $n$ levels. Hierarchical models can improve query performance as search can be conducted in an upwards manner only over the graph. This reduced the space complexity as edges are stored at their lower endpoints only.

Geisberger et al.~\cite{Geisberger2008} propose contraction hierarchies, where vertices are initially ordered by importance, and then a hierarchy is generated by contracting the least important vertices in an iterative manner. Contracting is the process of replacing the shortest-paths passing a vertex by what they call shortcuts. They propose a hierarchical algorithm that utilizes a bidirectional shortest-path search technique.

Batz et al.~\cite{Batz2010} propose a time-dependent version of the algorithm. It tackles time-dependent road networks where it proposes a fast and exact route planning algorithm. The issue it faces is space complexity. They tackle this problem by using approximations of piecewise-linear functions that lead to significant space reduction while preserving correctness. The proposed approach relies on approximating shortcuts and non-shortcuts to acquire  time-dependent edge weights. Then, these weights can then be used with their bidirectional search algorithm to create a corridor of shortcuts that can be searched.

Kieritzcite et al.~\cite{Kieritz2010} propose a distributed memory parallelization of time-dependent contraction hierarchies. The algorithm identifies vertices that can be contracted in every iteration. Parallelization is achieved when each process contracts its vertices independently and the vertices contractions do not overlap with each other. They attempt to approximate the ordering of the sequential algorithms used.

Geisberger et al.~\cite{Geisberger2012}  devise an algorithm based on contraction hierarchies to calculate continent-based shortest-paths. The preprocessing step relies on the hierarchical properties of road networks in order to add shortcut edges. They use a modified version of Dijkstra's algorithm that visits only a few hundred vertices that in turn makes it suitable to implement on mobile devices.

\subsubsection{Multi-Level Graphs}

In a multi-level overlay graph, if a set of vertices lie at a specific level, then the shortest-paths in that level do not use vertex from the upper levels. In turn, this method depends on the correct selection of vertices to act as landmarks on the higher levels.

Schulz et al.~\cite{Schulz2002} propose a multi-level graph-based decomposition method that targets space reduction. This method precomputed the shortest-paths and replaces the weights of single edges with a weight equal to the shortest-path length. The result is a subgraph that is smaller in size when compared with the original graph. The subgraph distances between a set of vertices is the same as the shortest-path graph distance between the same set of vertices in the original graph.

Holzer et al.~\cite{Holzer2009} introduce several vertex selection criteria on overlay graphs. These include criteria to determine a representative subset of the original graph. They investigate the criteria's effectiveness over multilevel overlay graphs and the speed-up achieved for shortest-path computation.

\subsubsection{Transit vertex Routing}

Transit vertex routing precomputed the shortest paths to and from all landmarks identified in a graph. The algorithm requires extensive preprocessing but exhibits very fast query time as it requires a limited number of look-ups between landmarks located in different locations.

Bast et al.~\cite{Bast2007} propose transit vertex routing. They suggest that a vertical and horizontal sweep are sufficient to compute the set of transit vertices. They also illustrate some techniques to make the approach more space-efficient.

Arz et al.~\cite{Arz2013} propose a variant of contraction hierarchies that achieves an order of magnitude speeds up , similar to the time needed to find contraction hierarchies. They propose a graph-theoretical locality filter that does not affect the query time.

\subsubsection{Hub Labeling}

Modeling road networks as a low-dimensional graph is a method used for computing the shortest paths. One method used for such modeling is the process of~\textit{labeling}. Algorithms for labeling have been introduced in the distributed computing field~\cite{Gavoille2004,Thorup2005}. In the labeling preprocessing stage, each vertex $v$ is computed and assigned a~\textit{forward label} and a~\textit{reverse label}. The forward label encompasses a set of vertices $w$, where each vertex contains a computed distance $dist(v,w)$ from $v$. The reverse label consists of a set of vertices $u$, where each vertex contains a computed distance $dist(u,v)$ to $v$. These labels are later used in the query stage to determine the vertices that minimize the distance from source to destination. A label can be perceived as a set of \textit{hubs} that a vertex $v$ has a direct connection to. The labeling algorithm ensures that any two vertices have one hub in common when computing the shortest path.

Hub labeling starts by preprocessing the vertices, where, for each vertex $v$, it precomputes the distance to a set of landmarks $L(v)$ in the vertex label. The query algorithm is fast as long as the number of landmarks of the source and destination vertices is small. Storing the labels in a consecutive manner allows the algorithm to exhibit good locality.

Abraham and Delling~\cite{Abraham2011,Abraham2012} propose a labeling scheme that, given a vertex $s$ and $t$, it considers the sets of vertices visited by the forward contraction hierarchy from $s$ and the reverse contraction hierarchy of $t$. The contraction hierarchies algorithm computes for the shortest-path the intersection of the forward and reverse sets that contain the maximum-rank vertex.

Babenko et al.~\cite{Babenko2013} propose an approximation algorithm for producing small labels. Their main target is to reduce the size of the maximum hub-label. This reduction process leads to unbalanced solutions as vertices will have a skewed label sizes. They propose an approximation algorithm for the maximum label size that runs in  $O(log n)$. The proposed approach reduces the the hub-labeling problem to a set-covering problem.

Cohen et al.~\cite{Cohen2003} propose a data structure for storing the reachability label using a 2-hops cover of all the paths in a graph. Each vertex $v \epsilon V$ precomputes the Label $L_{in}$ and $L_{out} \subseteq V$ such that, for any pair $s$ and $t$, at least one vertex is in $L_{out}(s) \bigcap L_{in}(t)$. The distance labeling query finds the shortest-path from source $s$ to destination $t$ by finding the minimum distance from $(L_{out} (s), x)$ to $(x, L_{in} (t))$ for each label $x \epsilon (L_{out} (s) \bigcap L_{in}(t))$. The size of a label $L$ is not guaranteed and the polynomial preprocessing time is approximately $O(log n)$ for finding a 2-hop cover of the invariant paths whose size is larger than the set of all shortest-paths. 

Chang et al.~\cite{Chang2012} propose a multi-hop distance labeling with a size smaller than another 2-hop labeling approach~\cite{Cohen2003}. In the preprocessing phase, the algorithm stores a parent function $P$ that assigns the parent vertex to each vertex by avoiding the preprocessing of the all-pairs shortest-path. The proposed approach performs vertex separation on the graph $G$ that divides $G$ into multiple connected subgraphs. The graph is further decomposed into a minimal tree $T (I,F)$, where $I \subset V$ represents the set of vertices and $F$ is the set of edges. The approach uses the distance query to compute the minimum distance. The time complexity of query processing is $O(tw*h)$, where $tw$ represents the width and $h$ represents the height of the decomposed tree $T$.

\subsubsection{Highway Node Routing}

The motivation behind using highway node routing is that prominent vertices that overlap various shortest-paths will generate sparse overlay graphs. The result would be faster query processing and lower space overhead. 

Schultes and Sanders~\cite{Schultes2007} proposes a dynamic algorithm that is space-efficient and allows query time to be thousand times faster when compared to Dijkstra's algorithm. The choice of vertices is achieved by capitalizing on previous results in addition to using the required vertex sets defined by highway hierarchies algorithms. They simplify the complications of computation into the prepreprocessing step. This also leads to simplification of the query processing algorithm, especially the dynamic variants.

Abraham~\cite{Abraham2010} suggests that road networks do not necessarily have a significant highway dimension. The proposed algorithm relies on realizing balls of a specific radius $r$. For every $r > 0$, there exits a sparse set  $S_{r}$ where shortest-path of length more than $r$ will have a vertex from the set $S_{r}$. If every ball having radius $O(r)$ contains less number of vertices than $S_{r}$, then the set $S_{r}$ is sparse.

\section{Dynamic Shortest-Path Algorithms}

The main requirement of dynamic shortest-path algorithms is to  process updates and query operations efficiently in an online fashion. In the update operation, edges are inserted or deleted from the graph. In the query operation,  the distance between vertices is computed.

\textit{Fully dynamic algorithms} are those that can process insertions and deletions. \textit{Incremental algorithms} can process insert operations, but not delete operations. \textit{Decremental algorithms} can process delete operations, but not insert operations. This implies that incremental and decremental algorithms are \textit{partially dynamic}. The following section illustrates the algorithms that demonstrate the aforementioned differences.

\subsection{All-Pairs Shortest-Path (APSP)}

The all-pairs shortest-paths algorithms reports the distances between any two vertices in a graph. The algorithms attempt to answer distance queries between any two vertices while dynamically maintaining changes that can occur to the graph such as inserts, deletes, and updates.

Demetrescu and Italiano~\cite{Demetrescu2004} propose a fully dynamic algorithm over directed graphs for all-pairs shortest-paths with real-valued edge weights. Every edge can have a predefined number of values. Their algorithm achieves an amortized time complexity of $O(S · n^{2.5} log^3 n)$ for update operations while achieving an optimal worst-case for query processing time. The proposed algorithm for the update operation inserts or deletes a vertex in addition to all its possible edges. The algorithm also maintains a complete distance matrix between updates.

Thorup~\cite{Thorup2004a} improves over Demetrescu and Italiano~\cite{Demetrescu2004} by reducing the fully-dynamic graph problem to a smaller set of decremental problems. Thorup adopts the idea of a fully-dynamic minimum spanning tree by utilizing the efficiency of the decremental algorithm to solve the fully-dynamic all-pairs shortest-paths problem.

Bernstein~\cite{Bernstein2009} presents a $(2 + \epsilon)$-approximation algorithm for APSP over an undirected graph with positive edge weights. Bernstein's algorithm achieves an update time that is almost linear and a query time of $O(log logn)$. The proposed query algorithm is deterministic while the update procedure is randomized. The algorithm run-time behavior depends on the distance from the source vertex to the destination vertex. Since $d(x, y)$ is not known beforehand, the algorithm relies on guessing  several different values for $d(x, y)$.

Roditty and Zwick~\cite{Roditty2010} propose a fully dynamic APSP algorithm for unweighted directed graphs. The algorithm is randomized and the correctness of the returned results are claimed to be high. The proposed algorithm passes through a set of phases that rely on the ideas of a decremental algorithm~\cite{Henzinger1995}. They demonstrate how the incremental and decremental versions of the SSSP problems are similar in terms of complexity to the the static all-pairs shortest-paths problem over directed or undirected graphs

Bernstein~\cite{Bernstein2013} proposes an $(1 + \epsilon)$ approximate algorithm that improves over existing studies with respect to the delete operation and edge weight increase. The algorithm computes the decremental all-pairs shortest-paths on weighted graphs. The approach achieves an update time of $o(mn^2)$ using a randomized algorithm.

Henzinger et al.~\cite{Henzinger2013} enhances over the fastest deterministic algorithm by Shiloach and Even~\cite{Shiloach981} by achieving an update time of $Ȏ(n^{5/2})$. It also achieves a constant query time. Also, they propose a deterministic algorithm with with an update time of $Ȏ(mn)$ and a query time of $O(log log n)$. They introduce two techniques, namely a lazy Even-Shiloach tree algorithm. The proposed approach maintains a shortest-paths tree that is bounded by distance with a Even-Shiloach tree based de-randomization technique.

\subsection{Single-Source Shortest-Path}

The single-source shortest-paths algorithm reports the distances from a given source vertex. The dynamic algorithm computes the update and query operations in an online fashion. The update operation inserts, deletes, or modify the edge's weight. The query operation probes for the distance from the source vertex to a given target vertex.

Fakcharoenphol and Rao~\cite{Fakcharoenphol2006} propose an algorithm for planar graphs with real-valued edge weights. It achieves a time complexity of $O(nlog^3 n)$. It performs update and query operations in $O(n^{4/5} log^{13/5} n)$ amortized time. The proposed algorithm uses Monge matrices~\cite{Cechlarova1990} with a combination of Bellman-Ford and Dijkstra's algorithms for searching ~in sub-linear time. 

Bernstein and Roditty~\cite{Bernstein2011} propose a dynamic shortest-paths algorithm that can achieve an update time better than $O(n)$ without sacrificing query time. In specific, they obtain $O(n^{2+o(1)})$ total update time and constant query time. The main type of graphs that it can achieve this result on is  moderately sparse graphs. Bernstein and Roditty propose two randomized decremental algorithms that operate over unweighted, undirected graph for two approximate shortest-path problems.

Henzinger et al.~\cite{Henzinger2014} improve the update operation time of Bernstein and Roditty~\cite{Bernstein2011} to $O(n^{1.8+o(1)} + m^{1+o(1)})$ while maintaining a constant query time. The algorithm utilizes the center-cover data structure where, given a parameter $h$ and a constant $\gamma$, maintains $O(h)$ vertices, referred to as \textit{centers}. The main property of the center-cover data structure is that every vertex within a specific distance is in a tree termed Even-Shiloach tree (ES-tree). The proposed algorithm has the same property of the center-cover data structure and is fastest when $h$ is moderately small.

\section{Time-Dependent Shortest-Path Algorithms}

A time-dependent shortest-path algorithm processes graphs that have edges associated with a function, known as an \textit{edge-delay} function. The edge-delay function indicates how much time is needed to travel from one vertex to another vertex. The query operation probes for the the minimum-travel-time path from the source to the destination vertex over graph. The returned result represents the best departure time found in a given time interval.

\subsection{Continuous-Time Algorithms}

Kanoulas et al.~\cite{Kanoulas2006} propose an algorithm that finds a set of all fastest paths from source to destination given a specified time interval. The specified interval is defined by the user and represents the departure or arrival time. The query algorithm finds a partitioning scheme for the time interval and creates a set of sub-intervals where each sub-interval is assigned to a set of fastest paths. Unlike the $A^{*}$ algorithm, the proposed algorithm probes the graph only once instead of multiple times.

Ding et al.~\cite{Ding2008} propose an algorithm that finds the departure time that minimizes the travel time over a road network. Also, the traffic conditions are dynamically changing in the road network. The algorithm is capable of operating on a variety of time-dependent graphs.

George et al.~\cite{George2006,George2007} propose a Time-Aggregated Graph (TAG) graph that changes its topology with time. In TAG, vertices and edges are modeled as time series. Apart from time dependence, it is also responsible for managing the edges and vertices that are absent during any instance in time.  They propose two algorithms to compute shortest-path using time-aggregated network (SP-TAG) and best start-time shortest-path (BEST). SP-TAG finds the shortest-path at the time of given query using a greedy algorithm. On the other hand, BEST algorithm finds out the best start-time (i.e., earliest travel time) over the entire period using TAG. The time complexity of SP-TAG and BEST are $O(e(log T + log n)$, and $O(n^{2}eT)$, respectively, where $e$ represents edges, $n$ represents vertices, and $T$ represents the time instance.

Ding et al.~\cite{Ding2008} propose an algorithm for the shortest-path problem over a large time-dependent graph GT. Each edge has a delay function that denotes the time taken from the source vertex to the destination vertex at a given time. The user queries the least travel time (LTT). The proposed algorithm achieves a space complexity of $O((n + m)\alpha(T))$ and a time complexity of $O((n log n + m)\alpha(T))$.

\subsection{Discrete-Time Algorithms}

Nannicini et al.~\cite{Nannicini2008a} propose a bidirectional $A^{*}$ algorithm that restricts the $A^{*}$ search  to a set of vertices that are defined by a time-independent algorithm. The bidirectional~$A^{*}$~algorithm operates in two modes, where the first mode, namely the\textit{forward search} algorithm, is run on the graph weighted by a specific cost function while the second mode, namely the \textit{backward search}, is run on the graph weighted by a lower-bound function.

Delling and Wagner~\cite{Delling2009} reanalyzes various time-dependent technique. The concluded that the most of the techniques that operate over time-dependent graphs guarantee correctness by augmenting the preprocessing and query phases subroutines.

Foschini et al.~\cite{Foschini2014} study the computational complexity of the shortest-paths problem over time-dependent graphs. They conclude that linear edge-cost functions causes the shortest path to the destination changes $n^{\theta(logn)}$ times. They study the complexity of the arrival time by mapping the problem to a parametric shortest-paths problem in order for it to be analyzed correctly.

Demiryurek et al.~\cite{Demiryurek2011} propose a technique to speed-up the fastest-path computation over time-dependent spatial graphs. They propose a technique based on the $A^{*}$ bidirectional time-dependent algorithm that operates in two main stages. The first stage is pre-computation, where it partitions the graph into a set of partitions that do not overlap. Next, they calculate a lower-bound distance label for vertices and borders. The second state is online, where it probes for the fastest path by utilizing a heuristic function  based on the computed distance labels. The results indicate that the proposed technique decreases the computation time and reduces the storage complexity significantly.

\section{Stochastic Shortest-Path Algorithms}

A stochastic shortest-path attempts to capture the uncertainty associated with the edges by modeling them as random variables. Then, the objective becomes to compute the shortest-paths based on the minimum expected costs. The two notable lines of research in this problem are adaptive and non-adaptive algorithms. The adaptive algorithms determine what the next best next hop would be based on the current graph at a certain time instance. The non-adaptive algorithms focus on minimizing the length of the path.

\subsection{Adaptive Algorithms}

Miller-Hooks and Mahmassani~\cite{Miller-Hooks2000} propose an algorithm to determine the apriori least-expected-time-paths from all source vertices  to a single destination vertex. This computation is for done for each departure time during busy time of the graph. They also propose a lower-bound over these apriori least-expected-time-paths. 

Nikolova et al.~\cite{Nikolova2006} propose an algorithm that maximizes the probability without exceeding a specific threshold for the shortest-paths length. They define a probabilistic model where edge weights are drawn from a known probability distribution. The optimal path is the one with the maximum probability indicating a path that does not pass a specific threshold.

\subsection{Non-Adaptive Algorithms}

Loui~\cite{Loui1983} proposes using a utility function with the length of the path, where the utility function is monotone and non-decreasing. When the utility function exhibits a linear or an exponential behavior, it becomes separable into the edge lengths. This allows the utility function to be identified using classical shortest-paths algorithms via paths that maximize the utility function.

Nikolova et al.~\cite{Nikolova2006a} propose an algorithm for optimal route planning under uncertainty. They define the target as a function of both the path length and the departure time starting from the source. They indicate that path and start time are jointly optimizable due to the penalizing behavior that they exhibit for late and early arrivals. They also indicated that this joint optimization is reducible to classic shortest-path algorithms.

\section{Parametric Shortest-Path Algorithms}

Parametric shortest-paths objective is to compute the shortest-paths for all vertices based on a specific parameter. It probes for the parameter values known as \textit{breakpoints} where the shortest-path tends to change. The edge value varies based on a linear function of the parameter value.

Mulmuley and Shah~\cite{Mulmuley2001} propose a model for lower-bound computation. It is a variant of the Parallel Random Access Machine. The proof starts with a lower-bound definition about the parametric complexity of the shortest-path problem. Plotting the weights of the shortest-path as a function results in an optimal cost graph that is piecewise-linear and concave. Breakpoints are defined as a fixed set of linear weight functions over a fixed graph.

Young et al.~\cite{Young2002} propose a model where the computed edge values makes it more tractable than its predecessors. This tractability allows obtaining shortest-paths in polynomial time.  They use the algorithm proposed by Karp and Orlin~\cite{Karp1981} and modify it to use Fibonacci heaps instead in order to improve its performance.

Erickson~\cite{Erickson2010} proposes an algorithm for computing the maximum flow in planar graphs. The algorithm maintains three structures, namely an edge spanning tree, a predecessor dual vertex set, and the slack value of dual edge set. They compute the initial predecessor pointers and slacks in $O(nlogn)$ using Dijkstra's algorithm.

\section{Replacement Shortest-Path Algorithms}

Consider a Graph $G=(V,E)$, where $V$ is the set of vertices and $E$ is the set of edges. For every Edge $e~\varepsilon~E$ on the shortest-path from source $s~\varepsilon~V$ to destination $d~\varepsilon~V$, the replacement path algorithm calculates the shortest-path from $s$ to $d$ that avoids $e$.

Emek et al.~\cite{Emek2010} propose an algorithm that computes the replacement path in near-linear time. The algorithm requires $O(n log^3 n)$ time during the preprocessing stage and $O(h log log n)$ time to answer the replacement path query, where $h$ is the number of hops in a weighted planar directed graph.

Roditty and Zwick~\cite{Roditty2012} propose a Monte-Carlo randomized algorithm that computes the replacement path in an unweighted directed graph. The run-time complexity of the algorithm is $\tilde{O}(m \sqrt{n})$. The Monte Carlo algorithm improves the run-time of the $k$-simple shortest-path and Vickrey pricing problems~\cite{Hershberger2001} by a factor of $\sqrt{n}$.

Bernstein~\cite{Bernstein2010} proposes an approximate $(1+\epsilon)$ replacement-path algorithm that computes the paths in $O(\epsilon^{-1} log^{2}n(m + nlog(nC/c)(m+nlogn)) = \tilde{O} (m log (nC/c)/\epsilon)$ time, where $C/c$ is the ratio of largest and smallest edge-weights in the graph. Bernstein's algorithm achieves a running time of  $\tilde{O}(km\sqrt{n})$ when applied over the $k$-th simple shortest-paths problem.

\section{Alternative Shortest-Path Algorithms}

The alternative shortest-path problem reports paths that avoid a given vertex or edge, termed the unwanted vertex or the unwanted edge. The key difference between the replacement-path and the alternative shortest-path is that the user is not required to specify the unwanted vertex or edge for replacement paths. The goal of the alternative path problem is reusing the previously computed results of the unwanted vertex or edge. In turn, this achieves better performance. Existing algorithms, e.g., all-pairs dynamic shortest-paths, do not solve the alternative shortest-path problem because of the high complexity of the update operation.

Xie et al.~\cite{Xie2012} propose a storage schemed, termed iSPQF. It is an extension of  the shortest-path quad-tree~\cite{Sankaranarayanan2005} that further reduces the number of quad-trees at each vertex. The space complexity of the shortest-path quad-tree into forest (SPQF) is $O(n^{1.5})$. The SPQF algorithm can find the alternative shortest-path over a single source (from source $s$ to destination $d$ that avoids Vertex $v$) as well as all pairs (from set of sources $X$ to set of destinations $Y$ that avoid Vertex $v$) in $O(n)$ time-complexity.

\section{Weighted Region Shortest-Path Algorithms}

Mitchell and Papadimitriou~\cite{Mitchell1990} define the Weighted Region Problem (WRP) as a generalization of the two-dimensional shortest path problem with obstacles. The problem assumes that the plane is subdivided into weighted polygonal regions. The objective is to minimize the cost according to a weighted Euclidean metric. The study by Mitchell and Papadimitriou sheds light on the discriminating properties of the weighted region problem over planar divisions and proposes an algorithm that runs in $O(n^{8}L)$, where $n$ is the number of vertices and $L$ is the number of bits required to encode the problem instance. In specific, $L = O(log(nNW/ \epsilon W))$, where $N$ is the maximum integer representing vertices of the triangulation, and $\epsilon > 0$ is a user-specified error value that can be tolerated.

Mata and Mitchell~\cite{Mata1997} propose an algorithm to compute the approximate optimal-path for the weighted planar subdivision problem by constructing a sparse graph, termed the \textit{path-net}. The approach uses Snell's law of Refraction~\cite{Warntz1957}  to divide the vertices into cones that bound the path of a vertex. The worst-case complexity to build the path-net graph with $O(kn)$ vertices is $O(kn^{3})$, where $k$ is the number of cones. After being scanned, it produces the paths that are within a factor of $(1 + \epsilon)$ from the optimal solution.

\section{Conclusion}

In this paper, we devise a taxonomy for the shortest-path problem. For each branch of the taxonomy, we illustrate the discriminating features and highlight the state-of-the-art research. The taxonomy provides investigators of the shortest-path problem with a guideline on where a required problem definition maps within the current related work.

\section*{Acknowledgements}
Walid G. Aref's research has been supported in part by the National Science Foundation under Grant IIS 1117766.

\newcommand{\etalchar}[1]{$^{#1}$}

\end{document}